\documentclass[final,5p,fleqn]{elsarticle}
\usepackage{amsmath,amssymb}
\usepackage{epsfig}
\usepackage[dvips]{color}
\newcommand{\beq}{\begin{eqnarray}}
\newcommand{\eeq}{\end{eqnarray}}
\newcommand{\bed}{\mbox{$^{11}$Be+d}}
\newcommand{\bedp}{\mbox{$^{11}{\rm Be}(d,p)^{12}{\rm Be}$}}
\newcommand{\ben}{\mbox{$^{10}$Be+n}}
\newcommand{\bep}{\mbox{$^{10}$Be+p}}

\newcommand{\emax}{\mbox{$E_{\rm max}$}}

\journal{Physics Letters B}

\begin{document}
\begin{frontmatter}
	\title{Four-body continuum effects in $\bed$ scattering}
	\author[PNTPM]{P. Descouvemont\fnref{label2}}
	\ead{pdesc@ulb.ac.be}
\fntext[label2]{Directeur de Recherches FNRS}
\address[PNTPM]{Physique Nucl\'eaire Th\'eorique et Physique Math\'ematique, C.P. 229,
	Universit\'e Libre de Bruxelles (ULB), B 1050 Brussels, Belgium}

\begin{abstract}
We present a new reaction model, which permits the description of reactions where both colliding nuclei present 
a low threshold to breakup.  The method corresponds to a four-body extension of the Continuum Discretized Coupled
Channel (CDCC) model.  We first discuss the theoretical formalism, and then apply the method to $\bed$ 
scattering at $E_{\rm c.m.}=45.5$ MeV. The $^{11}$Be nucleus and the deuteron are described by $\ben$ and $p+n$
structures, respectively. The model involves very large bases, but we show that an accurate description of elastic-scattering data may be achieved only when continuum states
of $^{11}$Be and of the deuteron are introduced simultaneously.  We also discuss breakup calculations,
and show that the cross section is larger for $^{11}$Be than for the deuteron.  
The present theory provides reliable wave functions that may be used in the analysis of $(d,p)$ or $(d,n)$ experiments involving radioactive beams.
\end{abstract}

\end{frontmatter}

The study of exotic nuclei is one of the main interests in modern nuclear physics \cite{TSK13}. Owing to the 
radioactive nature of exotic nuclei, they cannot be used as targets, and extensive efforts have been 
made over the last few decades to achieve high-quality radioactive beams \cite{BNV13}.
Various processes, such as elastic scattering, breakup, fusion or nucleon transfer are used to 
derive properties of exotic nuclei \cite{CGD15}.  Many theoretical \cite{TJ99,MNJ09} and
 experimental \cite{CHR74,KGU10} works have been performed by using 
nucleon stripping in $(d,p)$ or $(d,n)$ reactions.  In these conditions, the radioactive
beam impinges a deuteron target. A nucleon (either a neutron or a proton) is transferred to the 
incident particle, and the other nucleon is detected.

On the theoretical side, reactions involving exotic nuclei have been considered by many authors, 
through a variety of approaches.  The main characteristic of exotic nuclei is 
their low breakup threshold, and reaction models should include continuum effects as accurately as possible. 
The first theoretical treatments of this process addressed the elastic scattering of deuterons on stable targets
\cite{Ra74}.
These showed that the low binding energy of the deuteron (2.2 MeV) indirectly modifies the elastic scattering cross section 
through breakup effects.  This property triggered the development of the Continuum Discretized Coupled Channel (CDCC) 
method, where the breakup of the projectile is simulated by a discrete approximation of the 
continuum \cite{Ra74,KYI86,AIK87,YMM12}.  The CDCC method is very successful in explaining deuteron 
scattering on various targets \cite{AM10}.
Concerning breakup, and more specifically deuteron breakup, many works have been performed. In particular, the adiabatic
approximation considers the $p-n$ coordinate as a parameter, and assumes that the deuteron energy remains
constant \cite{JS70,AMN81}. These approximations permit a strong simplification of the CDCC calculations, but are
accurate at high energies only. In addition, they cannot be directly applied to four-body systems.

The characteristics of deuteron breakup is also seen in weakly bound systems. As such, the deuteron may be considered as the simplest and lightest exotic nucleus. Moreover, it is clear that the CDCC method is well suited to 
reactions involving exotic nuclei, produced by radioactive ion-beam 
facilities.  A typical example is the $^{11}$Be nucleus where the neutron separation energy is 
0.50 MeV only.  However, as several nuclei present a three-body structure (such as $^6{\rm He}=\alpha +n+n$ or 
$^{11}{\rm Li}=^9{\rm Li}+n+n$), further development 
of the CDCC method has been made to deal with three-body projectiles \cite{MHO04}.  It has been clearly 
shown that both the three-body structure and the low breakup threshold must be included 
in the reaction model, to reproduce satisfactorily the experimental data.  More recently, extensions 
to microscopic approaches, where the projectile is described by a many-body structure, have been performed \cite{DH13}.

In current reaction models, one of the colliding nuclei (in general, the target) is assumed to 
be structureless.  While this approximation is quite justified for many reactions investigated so far,  
as mentioned before, an important contribution in the study of exotic nuclei is provided by the $(d,p)$ and 
$(d,n)$ stripping reactions (see, for example, Ref.\ \cite{KGU10} with $\bedp$). In these conditions,
the entrance channel is formed by two nuclei presenting a low breakup energy, and the traditional 
CDCC method, assuming that one of the colliding nuclei is structureless, 
is no longer sufficient.

In this Letter, we propose a new extension of the CDCC method, where the breakup of both nuclei is included,  
allowing the wave functions needed for nucleon-transfer reactions to be calculated.  
The present model can also be used to compute elastic scattering cross sections.  Recent data 
have been obtained for $\bed$ scattering at $26.9A$ MeV \cite{CLY16}.  The $\bed$ system provides an ideal 
test of our method: $^{11}$Be can be accurately described by a $^{10}$Be+n structure, and the $p + n$ 
structure of the deuteron is obvious.  The main goal of this work is to assess the importance 
of $^{11}$Be and $d$ breakup in the elastic process.  The applications, however, are not 
limited to elastic scattering.  The CDCC method provides other cross sections, such as inelastic, 
reaction or breakup cross sections.  Partial experimental data on $\bed$ breakup are
available \cite{CLY16} and will be compared to the present four-body model.

\begin{figure}[htb]
	\begin{center}
		\epsfig{file=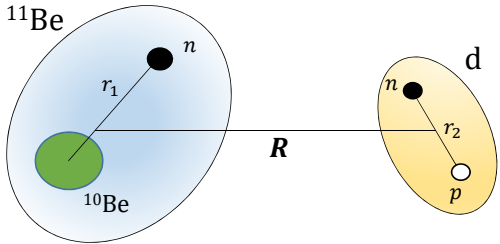,width=6.5cm}
		\caption{Cluster configuration and coordinates used in the four-body model for the $\bed$ system.}
		\label{fig_conf}
	\end{center}
\end{figure}

We consider the scattering of two nuclei, each of them presenting a two-body cluster structure.  The coordinates are shown in Fig.\ \ref{fig_conf} for the $\bed$ system: $\pmb{r}_1$ and $\pmb{r}_2$ are the internal coordinates, and
$\pmb{R}$ is the relative coordinate between the colliding nuclei.  The Hamiltonian of this four-body system is given by
\beq
H=H_1(\pmb{r}_1)+H_2(\pmb{r}_2)+T_R+V(\pmb{R},\pmb{r}_1,\pmb{r}_2),
\label{eq1}
\eeq
where $T_R$ is the relative kinetic energy. The potential term $V$ is defined from cluster-cluster optical potentials 
$U_{ij}$ as
\beq
V(\pmb{R},\pmb{r}_1,\pmb{r}_2)=\sum_{i=1}^2\sum_{j=1}^2 U_{ij},
\label{eq2}
\eeq
where the radial dependences of the $U_{ij}$ are easily expressed as a function of $\pmb{r}_1$, $\pmb{r}_2$,
and $\pmb{R}$. Notice that $U_{ij}$ contains the cluster-cluster Coulomb potential. In this way, Coulomb breakup
effects are included exactly.

In (\ref{eq1}), $H_1$ and $H_2$ are the internal Hamiltonians of the colliding nuclei, and are given by
\beq
H_i(\pmb{r}_i)=T_{\pmb{r}_i}+V_i(\pmb{r}_i),
\label{eq3}
\eeq
where $T_{\pmb{r}_i}$ is the internal kinetic energy, and $V_i(\pmb{r}_i)$ a two-body (real) potential 
describing nucleus $i$. The potentials are chosen so as to reproduce the low-lying states of the nucleus.

Our goal is to solve the Schr\"odinger equation associated with Eq.\ (\ref{eq1}) for scattering states.  This four-body scattering problem can be approximately solved with the CDCC.  We first define internal wave functions 
$\Phi_k^{I_i}(\pmb{r}_i)$ from 
\beq
H_i \Phi_k^{I_i}(\pmb{r}_i)=E_k^{I_i} \Phi_k^{I_i}(\pmb{r}_i),
\label{eq4}
\eeq
where $I_i$ is the angular momentum and $k$ the level of excitation (the internal parity is implied in $I_i$).  
In the CDCC method, the radial part of wave functions $\Phi_k^{I_i}(\pmb{r}_i)$ is expanded over a set of
 $N$ basis functions as
\beq
\Phi_k^{I_i}(r)=\sum_{n=1}^N c^{I_i k}_n \, u_n(r),
\label{eq4b}
\eeq
where $u_n(r)$ are appropriate functions, such as Gaussian or Lagrange functions.
In this way, Eq.\ (\ref{eq4}) 
is converted to a simple eigenvalue problem.  Negative energies $E_k^{I_i}$ correspond to physical states, 
and positive energies correspond to square-integrable approximations of the continuum \cite{KYI86}.  These states 
do not correspond to physical states, but are crucial to simulate the breakup of nuclei 1 and 2.  Many 
calculations have been performed within a three-body CDCC (see a recent review in Ref.\ \cite{YMM12}),  
but the four-body extension, which represents a huge increase in the computational demand, is necessary 
to investigate reactions where both nuclei have a low breakup threshold, such as $\bed$ for example.

The total four-body wave function is then expanded over the internal states as
\beq
\Psi^{JM\pi}_{\omega}(\pmb{R},\pmb{r}_1,\pmb{r}_2)=
\sum_c g^{J\pi}_{\omega,c}(R) \varphi^{JM\pi}_c(\Omega_R,\pmb{r}_1,\pmb{r}_2),
\label{eq5}
\eeq
where $\omega$ is the entrance channel, and where the channel function with orbital angular 
momentum $L$ is defined by
\beq
\varphi^{JM\pi}_c(\Omega_R,\pmb{r}_1,\pmb{r}_2)=
\biggl[ \bigl[ \Phi_{k_1}^{I_1}(\pmb{r}_1) \otimes \Phi_{k_2}^{I_2}(\pmb{r}_2)\bigr]^I 
\otimes Y_L(\Omega_R)\biggr] ^{JM}.
\label{eq6}
\eeq
In these definitions, $c$ indicates indices $c=(I_1,I_2,k_1,k_2,I,L)$, and $I$ is the channel spin. 
This coupling mode is standard in scattering theory. Let us 
discuss the summation over $c$ in Eq.\ (\ref{eq5}).  Continuum states of both nuclei are simulated by 
pseudostates $\Phi_k^{I_i}$ corresponding to positive energies $E_k^{I_i}$.  In actual applications, the summation is
truncated by a maximum energy and by a maximum angular momentum.  In practice, a reasonable number of 
pseudostates is in the range $\sim 30-50$.  In the present model, owing to the existence of pseudostates in both colliding 
nuclei, the number of channels greatly increases.  This leads to calculations where a thousands 
channels may be required.  Dealing with extremely large systems becomes feasible with modern computers, 
but still represents a challenge for the future of reaction models \cite{Th14}.

The radial functions $g^{J\pi}_{\omega,c}(R)$ are obtained from the coupled-channel system 
\beq
&&(T_L+E_c-E)g^{J\pi}_{\omega,c}(R)
+\sum_{c'}V^{J\pi}_{c,c'}(R) g^{J\pi}_{\omega,c'}(R)=0,
\label{eq7}
\eeq
with the kinetic-energy operator
\beq
T_L=-\frac{\hbar^2}{2\mu}\biggl(\frac{d^2}{dR^2}-\frac{L(L+1)}{R^2}  \biggr) .
\label{eq8}
\eeq
In these definitions, $\mu$ is the reduced mass, and $E_c$ is the energy of channel $c$.  The coupling 
potentials $V^{J\pi}_{c,c'}(R)$ are obtained from matrix elements of the potential (\ref{eq2}) between 
channel functions (\ref{eq6}).  In practice, potential (\ref{eq2}) is first expanded in multipoles, by 
numerical integration over the various angles (five angles).  Then the matrix elements $V^{J\pi}_{c,c'}(R)$ 
involve analytical integrals over the angles and numerical integrals over the radial coordinates $r_1$ and $r_2$.  
We use a Lagrange basis \cite{Ba15} to expand the radial functions (\ref{eq4}).  The main advantage 
of Lagrange functions is that integrals involving them are simple, and do not require 
any numerical quadrature (see, for example, Refs.\ \cite{DBD10,Ba15}).  

At large distance $R$, the 
radial functions tend to a combination of Coulomb functions as
\beq
g^{J\pi}_{\omega,c}(R)\rightarrow
v_{c}^{-1/2} \Bigl( I_{c} (k_{c}R)\delta_{c \omega} -  O_{c} (k_{c} R)U^{J\pi}_{\omega, c} \Bigr),
\label{eq9}
\eeq
where $I_c(x)$ and $O_c(x)$ are the incoming and outgoing Coulomb functions, and $k_c\ (v_c)$ is the wave 
number (velocity) in channel $c$.
Scattering states associated 
with (\ref{eq7}) are obtained within the $R$-matrix theory \cite{DB10,De16a} which provides 
the scattering matrix $U^{J\pi}_{\omega c}$.  From
scattering matrices in all partial waves $J\pi$, the various cross sections can be obtained by standard
formulae \cite{Sa83}.  

Recently obtained data, complemented by partial break\-up data, provide an opportunity to test the present four-body model. Chen {\sl et al}.\ \cite{CLY16} studied the $\bed$ elastic scattering and breakup
at $E_{\rm lab}(^{11}{\rm Be})=26.9A$ MeV \cite{CLY16}, which corresponds 
to $E_{c.m.}=45.5$ MeV. The $^{11}{\rm Be}$ nucleus is described by a $\ben$ potential \cite{CGB04}, including 
a spin-orbit term.  The $1/2^+$ and $1/2^-$ bound-state energies are adjusted by an appropriate choice 
of the potential.  For the deuteron, the $p + n$ Minnesota potential \cite{TLT77} is adopted; this nucleon-nucleon 
interaction fits the experimental deuteron ground state, and some low-energy scattering properties.  

The total potential $V$ [see Eq.\ (\ref{eq2})] involves four optical potentials: $\ben$ and $\bep$
are taken from the Koning-Delaroche global potential \cite{KD03}.  We choose the Minnesota potential 
for the $n + n$ and $n + p$ interactions.  To test the sensitivity of the cross sections against 
the optical potential, we also perform calculations with the Chapel Hill \cite{VTM91} compilation 
(referred to as CH89) for $\ben$ and $\bep$.  The CDCC calculations are performed with 
$I_1=1/2^{\pm},3/2^{\pm},5/2^+$ (i.e.\ orbital angular momenta $0,1,2$)
and $\emax=10$ MeV for $\ben$, and with $I_2=0^+,2^+$ and $\emax=15$ MeV for $p+n$.  
These two-body systems are described by 25 Gauss-Laguerre functions with a scaling parameter $h = 0.4$ fm 
(see Refs.\ \cite{Ba15,DBD10} for detail).  For the $\bed$ relative motion, we use angular momenta up 
to $J_{\rm max}=71/2$; and a channel radius $a = 25$ fm with 50 Gauss-Legendre basis functions.  Many tests have been performed to check the stability of the cross sections when these numerical conditions are varied.  In particular,
a special attention must be paid to the choice of the channel radius, which stems from a compromise \cite{DB10}.
Large values need many basis functions, and small values may not satisfy the $R$-matrix conditions. In
systems involving heavy targets, Coulomb couplings need in general values larger than 25 fm. However, for 
light systems, channel couplings around 25 fm are small enough to provide stable cross sections.

\begin{figure}[htb]
	\begin{center}
		\epsfig{file=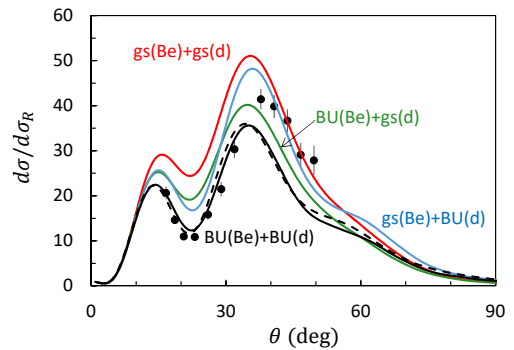,width=6.5cm}
		\caption{Ratio of the $\bed$ elastic cross section to the Rutherford cross section  at $E_{\rm c.m.}=45.5$ MeV. 
			Notation ``gs" means that only the ground state is included, and ``BU" that all continuum states
			are included.
			The black dashed line is obtained with the CH89 optical
			potential for $\ben$ and $\bep$. The data are reproduced from Ref.\ \cite{CLY16}.}
		\label{fig_ela}
	\end{center}
\end{figure}

The $\bed$ elastic cross section data are shown in Fig.\ \ref{fig_ela}, together with calculations made
under four different conditions.  
In the first calculation (referred to as ``gs(Be)+gs(d)"), only the ground states of $^{11}{\rm Be}$ and $d$ are included (in other words, all
breakup effects are absent).  Although the shape of the cross section is reasonably well reproduced, its 
amplitude is 
overestimated, reaching a factor of two near the minimum around $22^{\circ}$.  The introduction, either 
of $^{11}{\rm Be}$ breakup, or of the deuteron breakup, improves the agreement.  However at 
small angles $\theta \lesssim 35^{\circ}$, where the error bars are the smallest, the CDCC still 
overestimates the experimental data in these conditions.  The angular region $\theta \le 35^{\circ}$ is very well 
reproduced when continuum states of $^{11}{\rm Be}$ and of the deuteron are included {\sl simultaneously}.  
This result is consistent with the expectation: as $^{11}{\rm Be}$ and $d$ present both a low breakup 
threshold, including pseudostates in both nuclei is necessary to accurately reproduce experiment.  

At large angles, the full calculation is slightly lower than the data (by about $15\%$).  
This type of discrepancy is also found in other theoretical calculations \cite{CLY16}.  The same calculation, involving the $^{11}{\rm Be}$ and $d$ breakup, has been repeated with the CH89
$\ben$ and $\bep$ optical potentials (dashed line).  The difference with the Koning-Delaroche potential 
is marginal. It shows that the main issue is to include continuum states of the target and of the projectile.

The authors of Ref.\ \cite{CLY16} perform a three-body CDCC calculation, where $^{11}{\rm Be}$ is described
by a $^{10}{\rm Be}+n$ configuration, and where the deuteron is considered as point-like. This model includes $^{10}{\rm Be}$ excitation, but ignores the explicit treatment of the deuteron
breakup. This missing process is simulated by effective $n+d$ and $^{10}{\rm Be}+d$ optical potentials, and the resulting
$\bed$ cross sections present some sensitivity to these potentials. In contrast, our model only involves
nucleon-nucleus optical potentials, which are well known in the literature. It is possible that 
$^{10}{\rm Be}$ core excitations would have a weaker effect when
$^{11}{\rm Be}$ and $d$ breakups are explicitly taken into account. 

\begin{figure}[htb]
	\begin{center}
		\epsfig{file=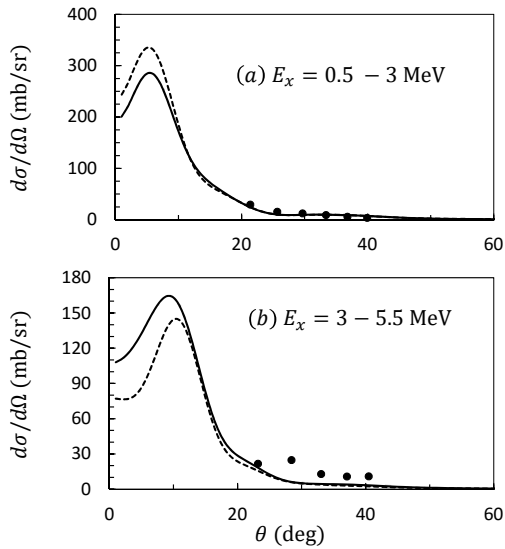,width=6.5cm}
		\caption{$\bed$ breakup cross sections at $E_{\rm c.m.}=45.5$ MeV, for two $^{11}$Be energy ranges. The dashed lines are obtained with the CH89 optical
			potential for $\ben$ and $\bep$. The data are taken from Ref.\ \cite{CLY16}.}
		\label{fig_bu1}
	\end{center}
\end{figure}

The present model also provides breakup cross sections.  In Ref.\ \cite{CLY16}, the $^{11}{\rm Be}$ breakup was 
measured by detecting events in two energy ranges, corresponding either to $E_x\sim 0.5-3$ MeV or to 
$E_x\sim3-5.5$ MeV. 
The CDCC results are presented in Fig.\ \ref{fig_bu1}, which shows an excellent agreement between theory 
and experiment in the range $E_x\sim 0.5-3$ MeV.  For higher $^{11}{\rm Be}$ energies, the calculations 
slightly underestimates the data above $\theta \approx 25^{\circ}$, but the angular dependence is realistic.  

Having data at smaller angles would be welcome since the breakup cross section is predicted
to be significantly larger.  Again, to test for systematic dependence on the choice of potential, 
we have performed calculations with the 
CH89 $\ben$ and $\bep$ optical potentials.  Differences appear at small angles $(\theta  < 10^{\circ})$, 
but both interactions provide similar cross sections in the experimental range.

\begin{figure}[htb]
	\begin{center}
		\epsfig{file=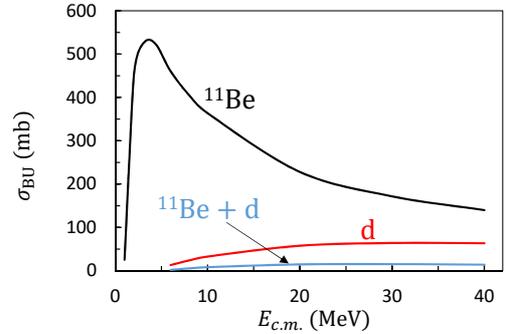,width=6.5cm}
		\caption{$\bed$ integrated breakup cross sections. The labels refer to single breakup ($^{11}$Be or $d$),
		and to double breakup ($^{11}{\rm Be}+d$).}
		\label{fig_bu2}
	\end{center}
\end{figure}

In Fig.\ \ref{fig_bu2}, we analyze the integrated breakup cross section up to $E_{c.m.}=40$ MeV.  The model 
provides single breakup (only $^{11}{\rm Be}$ or $d$ breaks up) and double breakup (both nuclei break up)
cross sections.  
The integrated cross sections are obtained from the scattering matrices as
\beq
\sigma_{\rm BU}=\frac{\pi}{2k_{\omega}^2}\sum_{J\pi}(2J+1)
\sum_c \vert U^{J\pi}_{\omega,c}\vert^2,
\label{eq10}
\eeq
and the summation over $c$ involves channels as required by the considered process.

The model predicts $^{11}{\rm Be}$ single breakup to be the dominant process.  
This difference between $^{11}{\rm Be}$ and $d$ breakup can be explained by the smallness of $E1$ 
contribution in $d$ breakup.  As expected, the second order process ($^{11}{\rm Be}$ and $d$ breakup) 
is small compared to the first order cross section.

In conclusion, we propose an extension of the CDCC method, where the breakup of both colliding nuclei is 
explicitly included.  The model is based on a coupled-channel approach, where the number of channels can be extremely 
large (up to several thousands).  However, this limitation can be solved with modern computer capabilities. 
The $R$-matrix method, used to solve the coupled-channel equation (\ref{eq7}) is stable, even for closed 
channels.   Using propagation methods \cite{DB10} is necessary with large bases, since they permit a
significant reduction of the computer times.

Application to  $\bed$ elastic scattering and comparison with recent data shows that the cross section 
without any breakup effect is
significantly overestimated. In contrast, the introduction of the $^{11}$Be and deuteron breakup provides an
excellent agreement with experiment. This result is obtained without any parameter fit, since the only
inputs are the optical potentials, which are taken from the literature. We have shown that the sensitivity to
the $\bep$ and $\ben$ potentials is weak. The main requirement to reproduce the data is to include
all breakup effects. The model is not limited to elastic scattering. As in other CDCC approaches, breakup or inelastic cross sections can also be derived. A possible improvement of the present model is to include antisymmetrization effects
between the neutrons associated with $d$ and with $^{11}$Be. However, this goes far beyond the CDCC limitation, i.e. that
the optical potentials between the constituents are local and angular-momentum independent.
 
Other applications may be considered, such as $^{8}$Li+d \cite{TMN11} or $^{7}$Li+$^{7}$Be 
\cite{BDJ05} for example.  
Also in nuclear astrophysics, reactions such as $^{13}$C+$^{13}$C \cite{ETJ11} need accurate models.  
On the other hand, 
DWBA analyses of $(d,p)$ and $(d,n)$ cross sections involve nucleus + $d$ wave functions \cite{Sa83}, which 
could be taken from the present model. A further extension to systems with two- and three-body nuclei, or even with
two three-body nuclei, is feasible but represents a computational challenge for future reaction modeling.

\section*{Acknowledgments}
I am grateful to Alex Murphy for his careful reading of the manuscript.
The present research is supported by the IAP programme P7/12 initiated by the Belgian-state 
Federal Services for Scientific, Technical and Cultural Affairs.
It benefited from computational resources made available on the Tier-1 supercomputer of the 
F\'ed\'eration Wallonie-Bruxelles, infrastructure funded by the Walloon Region under the grant agreement No. 1117545. 

\section*{References}

\begin{thebibliography}{31}
	\expandafter\ifx\csname natexlab\endcsname\relax\def\natexlab#1{#1}\fi
	\providecommand{\bibinfo}[2]{#2}
	\ifx\xfnm\relax \def\xfnm[#1]{\unskip,\space#1}\fi
	\bibitem[{Tanihata et~al.(2013)Tanihata, Savajols, and Kanungo}]{TSK13}
	\bibinfo{author}{I.~Tanihata}, \bibinfo{author}{H.~Savajols},
	\bibinfo{author}{R.~Kanungo}, \bibinfo{journal}{Prog. Part. Nucl. Phys.}
	\bibinfo{volume}{68} (\bibinfo{year}{2013}) \bibinfo{pages}{215}.
	\bibitem[{Blumenfeld et~al.(2013)Blumenfeld, Nilsson, and Duppen}]{BNV13}
	\bibinfo{author}{Y.~Blumenfeld}, \bibinfo{author}{T.~Nilsson},
	\bibinfo{author}{P.~V. Duppen}, \bibinfo{journal}{Physica Scripta}
	\bibinfo{volume}{2013} (\bibinfo{year}{2013}) \bibinfo{pages}{014023}.
	\bibitem[{Canto et~al.(2015)Canto, Gomes, Donangelo, Lubian, and
		Hussein}]{CGD15}
	\bibinfo{author}{L.~F. Canto}, \bibinfo{author}{P.~R.~S. Gomes},
	\bibinfo{author}{R.~Donangelo}, \bibinfo{author}{J.~Lubian},
	\bibinfo{author}{M.~S. Hussein}, \bibinfo{journal}{Phys. Rep.}
	\bibinfo{volume}{596} (\bibinfo{year}{2015}) \bibinfo{pages}{1}.
	\bibitem[{Timofeyuk and Johnson(1999)}]{TJ99}
	\bibinfo{author}{N.~K. Timofeyuk}, \bibinfo{author}{R.~C. Johnson},
	\bibinfo{journal}{Phys. Rev. C} \bibinfo{volume}{59} (\bibinfo{year}{1999})
	\bibinfo{pages}{1545}.
	\bibitem[{Moro et~al.(2009)Moro, Nunes, and Johnson}]{MNJ09}
	\bibinfo{author}{A.~M. Moro}, \bibinfo{author}{F.~M. Nunes},
	\bibinfo{author}{R.~C. Johnson}, \bibinfo{journal}{Phys. Rev. C}
	\bibinfo{volume}{80} (\bibinfo{year}{2009}) \bibinfo{pages}{064606}.
	\bibitem[{Cooper et~al.(1974)Cooper, Hornyak, and Roos}]{CHR74}
	\bibinfo{author}{M.~D. Cooper}, \bibinfo{author}{W.~F. Hornyak},
	\bibinfo{author}{P.~G. Roos}, \bibinfo{journal}{Nucl. Phys. A}
	\bibinfo{volume}{218} (\bibinfo{year}{1974}) \bibinfo{pages}{249}.
	\bibitem[{Kanungo et~al.(2010)Kanungo, Gallant, Uchida, Andreoiu, Austin,
		Bandyopadhyay, Ball, Becker, Boston, Boston, Brown, Buchmann, Colosimo,
		Clark, Cline, Cross, Dare, Davids, Drake, Djongolov, Finlay, Galinski,
		Garrett, Garnsworthy, Green, Grist, Hackman, Harkness, Hayes, Howell, Hurst,
		Jeppesen, Leach, Macchiavelli, Oxley, Pearson, Pietras, Phillips, Rigby,
		Ruiz, Ruprecht, Sarazin, Schumaker, Shotter, Sumitharachchi, Svensson,
		Tanihata, Triambak, Unsworth, Williams, Walden, Wong, and Wu}]{KGU10}
	\bibinfo{author}{R.~Kanungo}, \bibinfo{author}{A.~Gallant},
	\bibinfo{author}{M.~Uchida}, \bibinfo{author}{C.~Andreoiu},
	\bibinfo{author}{R.~Austin}, \bibinfo{author}{D.~Bandyopadhyay},
	\bibinfo{author}{G.~Ball}, \bibinfo{author}{J.~Becker},
	\bibinfo{author}{A.~Boston}, \bibinfo{author}{H.~Boston},
	\bibinfo{author}{B.~Brown}, \bibinfo{author}{L.~Buchmann},
	\bibinfo{author}{S.~Colosimo}, \bibinfo{author}{R.~Clark},
	\bibinfo{author}{D.~Cline}, \bibinfo{author}{D.~Cross},
	\bibinfo{author}{H.~Dare}, \bibinfo{author}{B.~Davids},
	\bibinfo{author}{T.~Drake}, \bibinfo{author}{M.~Djongolov},
	\bibinfo{author}{P.~Finlay}, \bibinfo{author}{N.~Galinski},
	\bibinfo{author}{P.~Garrett}, \bibinfo{author}{A.~Garnsworthy},
	\bibinfo{author}{K.~Green}, \bibinfo{author}{S.~Grist},
	\bibinfo{author}{G.~Hackman}, \bibinfo{author}{L.~Harkness},
	\bibinfo{author}{A.~Hayes}, \bibinfo{author}{D.~Howell},
	\bibinfo{author}{A.~Hurst}, \bibinfo{author}{H.~Jeppesen},
	\bibinfo{author}{K.~Leach}, \bibinfo{author}{A.~Macchiavelli},
	\bibinfo{author}{D.~Oxley}, \bibinfo{author}{C.~Pearson},
	\bibinfo{author}{B.~Pietras}, \bibinfo{author}{A.~Phillips},
	\bibinfo{author}{S.~Rigby}, \bibinfo{author}{C.~Ruiz},
	\bibinfo{author}{G.~Ruprecht}, \bibinfo{author}{F.~Sarazin},
	\bibinfo{author}{M.~Schumaker}, \bibinfo{author}{A.~Shotter},
	\bibinfo{author}{C.~Sumitharachchi}, \bibinfo{author}{C.~Svensson},
	\bibinfo{author}{I.~Tanihata}, \bibinfo{author}{S.~Triambak},
	\bibinfo{author}{C.~Unsworth}, \bibinfo{author}{S.~Williams},
	\bibinfo{author}{P.~Walden}, \bibinfo{author}{J.~Wong},
	\bibinfo{author}{C.~Wu}, \bibinfo{journal}{Phys. Lett. B}
	\bibinfo{volume}{682} (\bibinfo{year}{2010}) \bibinfo{pages}{391}.
	\bibitem[{Rawitscher(1974)}]{Ra74}
	\bibinfo{author}{G.~H. Rawitscher}, \bibinfo{journal}{Phys. Rev. C}
	\bibinfo{volume}{9} (\bibinfo{year}{1974}) \bibinfo{pages}{2210}.
	\bibitem[{Kamimura et~al.(1986)Kamimura, Yahiro, Iseri, Sakuragi, Kameyama, and
		Kawai}]{KYI86}
	\bibinfo{author}{M.~Kamimura}, \bibinfo{author}{M.~Yahiro},
	\bibinfo{author}{Y.~Iseri}, \bibinfo{author}{S.~Sakuragi},
	\bibinfo{author}{H.~Kameyama}, \bibinfo{author}{M.~Kawai},
	\bibinfo{journal}{Prog. Theor. Phys. Suppl.} \bibinfo{volume}{89}
	(\bibinfo{year}{1986}) \bibinfo{pages}{1}.
	\bibitem[{Austern et~al.(1987)Austern, Iseri, Kamimura, Kawai, Rawitscher, and
		Yahiro}]{AIK87}
	\bibinfo{author}{N.~Austern}, \bibinfo{author}{Y.~Iseri},
	\bibinfo{author}{M.~Kamimura}, \bibinfo{author}{M.~Kawai},
	\bibinfo{author}{G.~Rawitscher}, \bibinfo{author}{M.~Yahiro},
	\bibinfo{journal}{Phys. Rep.} \bibinfo{volume}{154} (\bibinfo{year}{1987})
	\bibinfo{pages}{125}.
	\bibitem[{Yahiro et~al.(2012)Yahiro, Matsumoto, Minomo, Sumi, and
		Watanabe}]{YMM12}
	\bibinfo{author}{M.~Yahiro}, \bibinfo{author}{T.~Matsumoto},
	\bibinfo{author}{K.~Minomo}, \bibinfo{author}{T.~Sumi},
	\bibinfo{author}{S.~Watanabe}, \bibinfo{journal}{Prog. Theor. Phys. Supp.}
	\bibinfo{volume}{196} (\bibinfo{year}{2012}) \bibinfo{pages}{87}.
	\bibitem[{Avrigeanu and Moro(2010)}]{AM10}
	\bibinfo{author}{M.~Avrigeanu}, \bibinfo{author}{A.~M. Moro},
	\bibinfo{journal}{Phys. Rev. C} \bibinfo{volume}{82} (\bibinfo{year}{2010})
	\bibinfo{pages}{037601}.
	\bibitem[{Johnson and Soper(1970)}]{JS70}
	\bibinfo{author}{R.~C. Johnson}, \bibinfo{author}{P.~J.~R. Soper},
	\bibinfo{journal}{Phys. Rev. C} \bibinfo{volume}{1} (\bibinfo{year}{1970})
	\bibinfo{pages}{976}.
	\bibitem[{Amakawa et~al.(1981)Amakawa, Mori, Nishioka, Yazaki, and
		Yamaji}]{AMN81}
	\bibinfo{author}{H.~Amakawa}, \bibinfo{author}{A.~Mori},
	\bibinfo{author}{H.~Nishioka}, \bibinfo{author}{K.~Yazaki},
	\bibinfo{author}{S.~Yamaji}, \bibinfo{journal}{Phys. Rev. C}
	\bibinfo{volume}{23} (\bibinfo{year}{1981}) \bibinfo{pages}{583}.
	\bibitem[{Matsumoto et~al.(2004)Matsumoto, Hiyama, Ogata, Iseri, Kamimura,
		Chiba, and Yahiro}]{MHO04}
	\bibinfo{author}{T.~Matsumoto}, \bibinfo{author}{E.~Hiyama},
	\bibinfo{author}{K.~Ogata}, \bibinfo{author}{Y.~Iseri},
	\bibinfo{author}{M.~Kamimura}, \bibinfo{author}{S.~Chiba},
	\bibinfo{author}{M.~Yahiro}, \bibinfo{journal}{Phys. Rev. C}
	\bibinfo{volume}{70} (\bibinfo{year}{2004}) \bibinfo{pages}{061601}.
	\bibitem[{Descouvemont and Hussein(2013)}]{DH13}
	\bibinfo{author}{P.~Descouvemont}, \bibinfo{author}{M.~S. Hussein},
	\bibinfo{journal}{Phys. Rev. Lett.} \bibinfo{volume}{111}
	(\bibinfo{year}{2013}) \bibinfo{pages}{082701}.
	\bibitem[{Chen et~al.(2016)Chen, Lou, Ye, Rangel, Moro, Pang, Li, Ge, Li, Li,
		Jiang, Sun, Zang, Zhang, Aoi, Ideguchi, Ong, Lee, Wu, Liu, Wen, Ayyad,
		Hatanaka, Tran, Yamamoto, Tanaka, Suzuki, and Nguyen}]{CLY16}
	\bibinfo{author}{J.~Chen}, \bibinfo{author}{J.~L. Lou}, \bibinfo{author}{Y.~L.
		Ye}, \bibinfo{author}{J.~Rangel}, \bibinfo{author}{A.~M. Moro},
	\bibinfo{author}{D.~Y. Pang}, \bibinfo{author}{Z.~H. Li},
	\bibinfo{author}{Y.~C. Ge}, \bibinfo{author}{Q.~T. Li},
	\bibinfo{author}{J.~Li}, \bibinfo{author}{W.~Jiang}, \bibinfo{author}{Y.~L.
		Sun}, \bibinfo{author}{H.~L. Zang}, \bibinfo{author}{Y.~Zhang},
	\bibinfo{author}{N.~Aoi}, \bibinfo{author}{E.~Ideguchi},
	\bibinfo{author}{H.~J. Ong}, \bibinfo{author}{J.~Lee},
	\bibinfo{author}{J.~Wu}, \bibinfo{author}{H.~N. Liu},
	\bibinfo{author}{C.~Wen}, \bibinfo{author}{Y.~Ayyad},
	\bibinfo{author}{K.~Hatanaka}, \bibinfo{author}{T.~D. Tran},
	\bibinfo{author}{T.~Yamamoto}, \bibinfo{author}{M.~Tanaka},
	\bibinfo{author}{T.~Suzuki}, \bibinfo{author}{T.~T. Nguyen},
	\bibinfo{journal}{Phys. Rev. C} \bibinfo{volume}{94} (\bibinfo{year}{2016})
	\bibinfo{pages}{064620}.
	\bibitem[{Thompson(2014)}]{Th14}
	\bibinfo{author}{I.~J. Thompson}, \bibinfo{journal}{J. Phys. G}
	\bibinfo{volume}{41} (\bibinfo{year}{2014}) \bibinfo{pages}{094009}.
	\bibitem[{Baye(2015)}]{Ba15}
	\bibinfo{author}{D.~Baye}, \bibinfo{journal}{Phys. Rep.} \bibinfo{volume}{565}
	(\bibinfo{year}{2015}) \bibinfo{pages}{1}.
	\bibitem[{Druet et~al.(2010)Druet, Baye, Descouvemont, and Sparenberg}]{DBD10}
	\bibinfo{author}{T.~Druet}, \bibinfo{author}{D.~Baye},
	\bibinfo{author}{P.~Descouvemont}, \bibinfo{author}{J.-M. Sparenberg},
	\bibinfo{journal}{Nucl. Phys. A} \bibinfo{volume}{845} (\bibinfo{year}{2010})
	\bibinfo{pages}{88}.
	\bibitem[{Descouvemont and Baye(2010)}]{DB10}
	\bibinfo{author}{P.~Descouvemont}, \bibinfo{author}{D.~Baye},
	\bibinfo{journal}{Rep. Prog. Phys.} \bibinfo{volume}{73}
	(\bibinfo{year}{2010}) \bibinfo{pages}{036301}.
	\bibitem[{Descouvemont(2016)}]{De16a}
	\bibinfo{author}{P.~Descouvemont}, \bibinfo{journal}{Comput. Phys. Commun.}
	\bibinfo{volume}{200} (\bibinfo{year}{2016}) \bibinfo{pages}{199}.
	\bibitem[{Satchler(1983)}]{Sa83}
	\bibinfo{author}{G.~R. Satchler}, \bibinfo{title}{Direct Nuclear Reactions},
	\bibinfo{publisher}{Oxford University Press}, \bibinfo{year}{1983}.
	\bibitem[{Capel et~al.(2004)Capel, Goldstein, and Baye}]{CGB04}
	\bibinfo{author}{P.~Capel}, \bibinfo{author}{G.~Goldstein},
	\bibinfo{author}{D.~Baye}, \bibinfo{journal}{Phys. Rev. C}
	\bibinfo{volume}{70} (\bibinfo{year}{2004}) \bibinfo{pages}{064605}.
	\bibitem[{Thompson et~al.(1977)Thompson, LeMere, and Tang}]{TLT77}
	\bibinfo{author}{D.~R. Thompson}, \bibinfo{author}{M.~LeMere},
	\bibinfo{author}{Y.~C. Tang}, \bibinfo{journal}{Nucl. Phys. A}
	\bibinfo{volume}{286} (\bibinfo{year}{1977}) \bibinfo{pages}{53}.
	\bibitem[{Koning and Delaroche(2003)}]{KD03}
	\bibinfo{author}{A.~J. Koning}, \bibinfo{author}{J.~P. Delaroche},
	\bibinfo{journal}{Nucl. Phys. A} \bibinfo{volume}{713} (\bibinfo{year}{2003})
	\bibinfo{pages}{231}.
	\bibitem[{Varner et~al.(1991)Varner, Thompson, McAbee, Ludwig, and
		Clegg}]{VTM91}
	\bibinfo{author}{R.~L. Varner}, \bibinfo{author}{W.~J. Thompson},
	\bibinfo{author}{T.~L. McAbee}, \bibinfo{author}{E.~J. Ludwig},
	\bibinfo{author}{T.~B. Clegg}, \bibinfo{journal}{Phys. Rep.}
	\bibinfo{volume}{201} (\bibinfo{year}{1991}) \bibinfo{pages}{57}.
	\bibitem[{Tengborn et~al.(2011)Tengborn, Moro, Nilsson, Alcorta, Borge,
		Cederk\"all, Diget, Fraile, Fynbo, Gomez-Camacho, Jeppesen, Johansson,
		Jonson, Kirsebom, Knudsen, Madurga, Nyman, Richter, Riisager, Schrieder,
		Tengblad, Timofeyuk, Turrion, Voulot, and Wenander}]{TMN11}
	\bibinfo{author}{E.~Tengborn}, \bibinfo{author}{A.~M. Moro},
	\bibinfo{author}{T.~Nilsson}, \bibinfo{author}{M.~Alcorta},
	\bibinfo{author}{M.~J.~G. Borge}, \bibinfo{author}{J.~Cederk\"all},
	\bibinfo{author}{C.~Diget}, \bibinfo{author}{L.~M. Fraile},
	\bibinfo{author}{H.~O.~U. Fynbo}, \bibinfo{author}{J.~Gomez-Camacho},
	\bibinfo{author}{H.~B. Jeppesen}, \bibinfo{author}{H.~T. Johansson},
	\bibinfo{author}{B.~Jonson}, \bibinfo{author}{O.~S. Kirsebom},
	\bibinfo{author}{H.~H. Knudsen}, \bibinfo{author}{M.~Madurga},
	\bibinfo{author}{G.~Nyman}, \bibinfo{author}{A.~Richter},
	\bibinfo{author}{K.~Riisager}, \bibinfo{author}{G.~Schrieder},
	\bibinfo{author}{O.~Tengblad}, \bibinfo{author}{N.~Timofeyuk},
	\bibinfo{author}{M.~Turrion}, \bibinfo{author}{D.~Voulot},
	\bibinfo{author}{F.~Wenander}, \bibinfo{journal}{Phys. Rev. C}
	\bibinfo{volume}{84} (\bibinfo{year}{2011}) \bibinfo{pages}{064616}.
	\bibitem[{Barua et~al.(2005)Barua, Das, Jhingan, Madhavan, Varughese, Sugathan,
		Kalita, Verma, Bhattacharjee, Datta, and Boruah}]{BDJ05}
	\bibinfo{author}{S.~Barua}, \bibinfo{author}{J.~J. Das},
	\bibinfo{author}{A.~Jhingan}, \bibinfo{author}{N.~Madhavan},
	\bibinfo{author}{T.~Varughese}, \bibinfo{author}{P.~Sugathan},
	\bibinfo{author}{K.~Kalita}, \bibinfo{author}{S.~Verma},
	\bibinfo{author}{B.~Bhattacharjee}, \bibinfo{author}{S.~K. Datta},
	\bibinfo{author}{K.~Boruah}, \bibinfo{journal}{Phys. Rev. C}
	\bibinfo{volume}{72} (\bibinfo{year}{2005}) \bibinfo{pages}{044602}.
	\bibitem[{Esbensen et~al.(2011)Esbensen, Tang, and Jiang}]{ETJ11}
	\bibinfo{author}{H.~Esbensen}, \bibinfo{author}{X.~Tang},
	\bibinfo{author}{C.~L. Jiang}, \bibinfo{journal}{Phys. Rev. C}
	\bibinfo{volume}{84} (\bibinfo{year}{2011}) \bibinfo{pages}{064613}.
	
\end{thebibliography}

\end{document}